\documentclass[preprint,pre,showpacs]{revtex4}
\usepackage{graphicx}
\usepackage{amssymb}
\usepackage{amsmath}
\newcommand{\nuperp}{\nu_\perp}
\begin{document}

\title{Critical behavior of the contact process in a multiscale network}

\author{Silvio C. Ferreira} \email{silviojr@ufv.br}
\author{Marcelo L. Martins} \email{mmartins@ufv.br}
\affiliation{Departamento de F\'{\i}sica, Universidade Federal Vi\c{c}osa, 36571-000, Vi\c{c}osa, MG, Brazil}

\date{\today}

\begin{abstract}
Inspired by dengue and yellow fever epidemics, we investigated the contact process (CP) in a multiscale network constituted by one-dimensional chains connected through a Barab\'asi-Albert scale-free network. In addition to the CP dynamics inside the chains, the exchange of individuals between connected chains (travels) occurs at a constant rate. A finite epidemic threshold and an epidemic mean lifetime diverging exponentially in the subcritical phase, concomitantly with a power law divergence of the outbreak's duration, were found. A generalized scaling function involving both regular and SF components was proposed for the quasistationary analysis and the associated critical exponents determined, demonstrating that the CP on this hybrid network and nonvanishing travel rates establishes a new universality class.
\end{abstract}

%\keywords{}
\pacs{05.70.Ln, 89.75.Da, 89.75.Hc, 05.70.Jk}

\maketitle
\parskip0cm

\section{Introduction}

Complex networks are involved in a wide range of biological, social and technological systems\cite{Albert1,Newman} such as, for instance, the Internet \cite{Albert1,Caldarelli}, the network of human sexual contacts \cite{Liljeros,Lind}, and the transportation infrastructure \cite{Gastner,Guimera}. These nets exhibit ``scale-free'' (SF) degree distributions, characterized by a probability $P(k)\sim k^{-\gamma}$ that an element in the network is connected to $k$ other elements. Usually, the degree exponent assumes a value in the range $2<\gamma<3$ \cite{Albert1}. In special, since many infectious diseases exploit the social contact network in order to spread among human hosts \cite{Barrett} and computer viruses inflict significant damages throughout the Internet \cite{Satorras1}, epidemic spreading through a variety of networks, including random graphs (RG), small worlds (SW) and SF networks, have been extensively studied \cite{Satorras1,Moore,Satorras2,Barthelemy,Castellano,Boguna,Serrano}.

Pastor-Satorras and coworkers \cite{Satorras1,Satorras2,Barthelemy,Boguna} investigated the epidemic spreading on complex networks using the susceptible-infected-susceptible (SIS) model. For all networks considered (RG, SW, and SF), excellent agreements between mean-field and numerical results were obtained at the transition to the absorbing state \cite{Satorras2}. However, Castellano and Pastor-Satorras \cite{Castellano} observed a non-mean-field critical behavior dependent on $\gamma$ for the classical Harris contact process (CP) \cite{Marro} on SF networks. So, these models, sharing the directed percolation (DP) universality class on regular lattices \cite{Marro,Boccara}, exhibit distinct critical behaviors on SF networks.

The deep impact of complex networks on epidemiology, radically altering traditional control and vaccination strategies, bring to the focus some epidemic processes associated to tropical diseases that can not be properly described by either a purely regular, SW or SF networks. Indeed, the spreading of dengue and yellow fever, the main re-emergent diseases \cite{note} in the world, are determined by the population dynamics of mosquitoes with a limited flight range ($\sim 200$ m for \textit{Aedes aegypti}), domestic habits and local dispersion independent on the human contact network. Furthermore, in their epidemiological chains, a relevant factor for viral circulation is the travels of symptomless infected people. Such travels occur on  regional and national interurban commuting traffic networks which behave as SF weighted nets with modularity \cite{Guimera}. Hence, the spreading of dengue and yellow fever combines a local contact process among mosquitoes and individuals around their houses with long-range dispersal through symptomless infected people traveling between cities. Aiming to include this two-component spreading process, Silva {\it et al.} studied the SIS model in SF networks of square lattices \cite{Silva}. Although they focused on dynamics instead of the transition to the absorbing state, a nonzero epidemic threshold characterized by a singular approach to the epidemic-free state was found.

\begin{figure*}[hbt]
\begin{center}
\includegraphics[width=9cm,height=!]{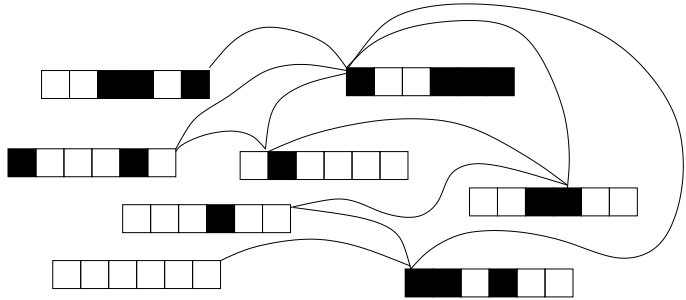}
\caption{\label{fig:network} Schematic representation of the network. The connections between chains (the cities) are represented by lines. Black and white squares correspond to occupied and empty sites, respectively.}
\end{center}
%\vspace{-.5cm}
\end{figure*}

In this paper, we report on numerical simulations of the CP model in a SF network in which the nodes are linear chains. The transition to the absorbing state is focused. The model and its computer implementation are presented in Sec. \ref{sec:model}. Simulations are reported and discussed in Sec. \ref{sec:result}. Finaly, some conclusions are drawn in Sec. \ref{sec:conclusion}

\section{Model}
\label{sec:model}

In the model, individuals lie on $N$ linear chains of size $L$ with periodic boundary conditions (the cities), in which empty and occupied sites represent healthy and infected individuals, respectively. The cities are connected according to the Barab\'asi-Albert (BA) model \cite{Albert1}, i. e., in a SF network with a degree distribution given by $P(k)\sim k^{-3}$ (FIG. \ref{fig:network}). The model dynamics incorporates a local spreading in which healthy sites with $n$ occupied nearest neighbors (NNs) are infected at a rate $n\lambda/2$ while any infected site is spontaneously cured at rate 1. Additionally, two sites in distinct connected nodes are interchanged at a rate $\alpha$ representing a travel. These travels occur preferentially to nodes with larger connectivity. The simulations were implemented as follows. At each time step, an infected site is selected at random and the time incremented by $\Delta t = 1/N_{occ}$, where $N_{occ}$ is the total number of infected sites in all nodes. The selected site will perform one of three actions: (i) become healthy with probability $p=1/(1+\lambda+\alpha)$; (ii) randomly select and infect one of its healthy NNs with probability $q=\lambda/(1+\lambda+\alpha)$; (iii) travel with probability $r=1-p-q$. In a travel, the target node $j$, with degree $k_j$, is chosen among all those connected to the departure node $i$ with probability $\Pi_{i \rightarrow j} =k_j/\sum_j k_j$. Notice that, differently from the Castellano and Pastor-Satorras model \cite{Castellano}, the infection spreading occurs among individuals with the same number of connections (the regular lattice coordination) as in the original CP \cite{Marro}, while in ref. \cite{Castellano} the number of connection varies from one individual to other following a power law distribution. Also,  the original CP is obtained when $\alpha = 0$ (no travels), but none limit corresponds to the CP on scale-free networks.

Computer simulations were done for chains of sizes varying from $L=10^2$ to $12800$ and number of nodes ranging from $N=400$ to $12800$. For each group of $10$ samples, a SF network was generated from $N_0=10$ fully interconnected nodes accordingly the BA preferential attachment algorithm~\cite{Albert1}. New nodes were sequentially added to the growing network through $m=4$ links, the minimum degree value. In order to determine the criticality at the transition to the absorbing state, the \textit{overall density} $\rho$ of infected sites  through all nodes, the \textit{local density} $\zeta$ restricted to those nodes with at least one infected site, the \textit{fraction of colonized nodes} $\Omega$ (those with infected sites), and the \textit{survival probability} $P_s$ were evaluated.

\section{Results}
\label{sec:result}

\begin{figure*}[hbt]
\begin{center}
\includegraphics[width=9cm,height=!,clip=true]{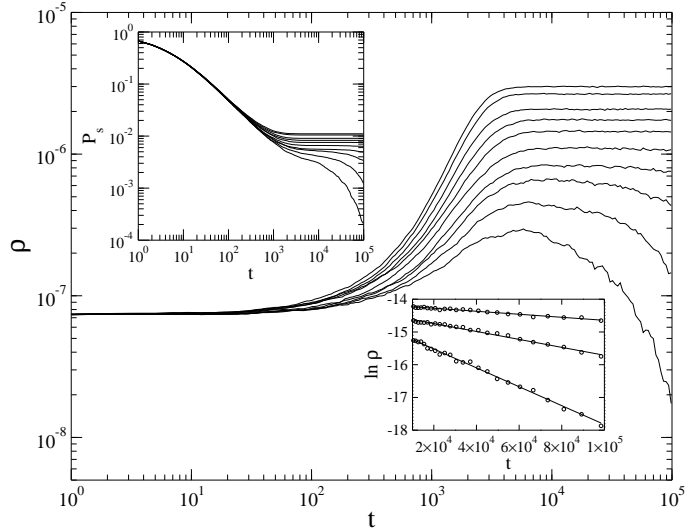}
\caption{\label{fig:rhops} Evolution in time of the density of infected sites for $L=10^4$, $N=1600$, $\alpha=0.5$, and epidemic rates varying from $\lambda=1.549$ to $1.558$ in intervals $\Delta \lambda=10^{-3}$ from the bottom to the top. Upper inset: corresponding survival probabilities as functions of time. Lower inset: semi-log plots for $\lambda=1.549 - 1.551$. The averages were done over $N_s=10^4$ to $5\times 10^5$ independent samples (the lower $\lambda$ the larger $N_s$).}
\end{center}
%\vspace{-.5cm}
\end{figure*}

As observed for the CP in SF networks \cite{Castellano}, the simulations in FIG. \ref{fig:rhops} revealed a non null epidemic threshold. The upper inset shows the corresponding survival probabilities, exhibiting the same behavior. The initial condition was a single infected site at the center of a randomly chosen node. In the upper-critical regime, both quantities reach constant asymptotic values, contrasting with the increasing as $n\sim t^d$ of the mean number of infected sites for the CP on regular lattices \cite{Marro}. The heterogeneous distributions of infected sites in the nodes, with only few of them highly infected, produce the plateaus observed at relatively short times ($\sim 10^3$ for chains with $L=6000$ sites). In general, models with absorbing configurations exhibit power law time dependencies $P_s\sim t^{-\delta}$, $n\sim t^\eta$ and $R^2\sim t^z$ at the critical point for the survival probability, the mean number of infected sites, and the spreading of epidemics, respectively \cite{Marro, r2,Jensen}. Accordingly, FIG. \ref{fig:rhops} provides spreading exponents close to zero for the CP model on the present hybrid network.

\begin{figure*}[hbt]
\begin{center}
\includegraphics[width=9cm,height=!,clip=true]{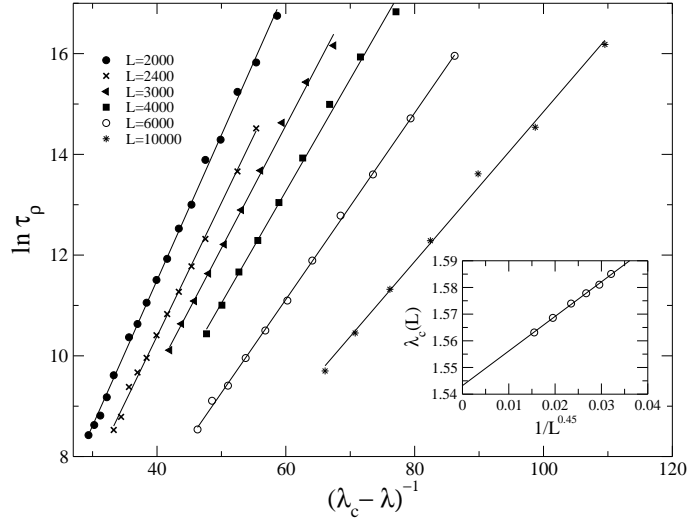}
\caption{\label{fig:txnt} Characteristic infection lifetimes in the sub-critical regime against the infection rates for several node sizes. $N=1600$ and $\alpha=0.5$ are fixed. Inset: FSS analysis to extrapolate the critical rate. Sampling as in FIG. \ref{fig:rhops}.}
\end{center}
%\vspace{-.5cm}
\end{figure*}

The mean lifetime divergence when $\lambda\rightarrow\lambda_c^-$ can be used to estimate the critical point. The exponential decay $\rho\sim\exp(-t/\tau_\rho)$ [lower inset of FIG. \ref{fig:rhops}] determines the decay time as a function of $\lambda$. As shown in FIG. \ref{fig:txnt}, an exponential divergence given by
$\tau_\rho(\lambda, L)\sim \exp(\mbox{const.}/\Delta)$ was found. Here, $\Delta=|\lambda_c-\lambda|$. This behavior differs from the power law $\tau\sim|\Delta|^{-\nu_\parallel}$ usually observed in transitions to absorbing states in regular lattices \cite{Marro}. Indeed, a power law also fits the data, but very large exponent values ($\nu_{\parallel}\approx 6$ to $8$) were obtained, supporting further the hypothesis of exponential divergence. The exponents $\nu_\parallel = \infty$ and $\delta = \eta = z = 0$ are consistent with the well known scaling relations $\delta = \beta / \nu_\parallel$, $z = 2\nu_\perp / \nu_\parallel$ and $4\delta+2\eta = d z$ \cite{Marro}. Despite very strong finite size effects, a characteristic feature of epidemics on SF networks \cite{Satorras1,Satorras2,Castellano}, the critical rate can be determined by exponential nonlinear fittings. In the inset of FIG. \ref{fig:txnt}, the critical rate is shown as a function of $L^{-0.45}$\cite{note1} and its extrapolation to an infinite size provides a critical rate $\lambda_c=1.5432(5)$, in which the error is in parenthesis. This value agrees with estimates from quasi-stationary simulations described in the next paragraphs. The critical rate and exponential divergence of $\tau_{\rho}$ does not depend on network size $N$ for the studied range. The same analysis was done for $\zeta$ and $P_s$ providing the same critical rate and exponential divergence for the mean lifetime estimated through the analysis of $\rho$.

\begin{figure*}[ht]
\begin{center}
\includegraphics[width=9cm,height=!,clip=true]{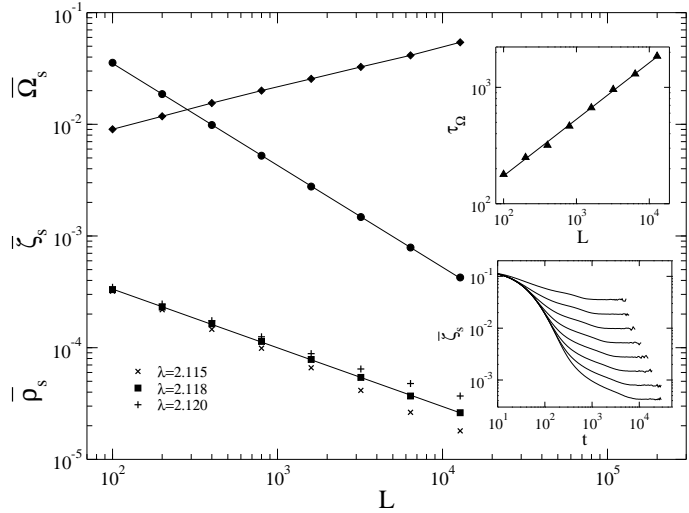}
\caption{\label{fig:rhoqs} QS densities $\overline{\rho}_s$ (squares), $\overline{\zeta}_s$ (circles) and $\overline{\Omega}_s$ (diamonds), for fixed $N=1600$ and $\alpha=0.1$ at the critical rate  $\lambda_c =2.118$. Upper- and a sub-critical data for $\rho$ are shown. Straight lines are power law fits. Lower inset: QS local densities as functions of time. Upper inset: Critical QS relaxation time determined using  the density $\overline{\Omega}_s$. The number of samples varies from $N_s=2\times10^3$ to $10^5$ (the larger $L$ the lower $N_s$).}
\end{center}
%\vspace{-.5cm}
\end{figure*}

In order to perform a finite size scaling (FSS) analysis near the criticality, we used quasistationary (QS) simulations in which only the survival trials are considered in the averages \cite{Marro,Castellano}. Moreover, a generalized scaling hypothesis that explicitly considers the validity of the usual FSS \cite{Marro, Castellano} for fixed $N$ or fixed $L$ was assumed

\begin{equation}
\label{eq:fss} 
\overline{\rho}_s(\Delta,L,N)=L^{-\beta/\nu_\perp} N^{-\beta^{\prime}/\nu_\perp^{\prime}} \mathcal{F}[\Delta L^{1/\nu_\perp} N^{1/\nu_\perp^{\prime}}].
\end{equation}
Similar relations can be proposed for the other quantities.  Thus, the thermodynamic limit can be estimated through three distinct ways. The first one, in which the size $L$ grows faster than the number of nodes $N$, implying $N/L\rightarrow 0 $, seems to be more adequate to describe epidemic spreading in the real world. Alternatively, the FSS analysis can be performed with $N$ growing faster than $L$, leading to $N/L \rightarrow \infty$, or maintaining a constant ratio $N/L$. 

For a large but fixed $N$, Eq. (\ref{eq:fss}) becomes $\overline{\rho}_s(\Delta,L)=L^{-\beta/\nu_\perp} \mathcal{G}(\Delta L^{1/\nu_\perp})$, in which $\mathcal{G}(x)\sim x^\beta$ for $x\gg 1$ and $\mathcal{G}(x)\sim x^{-\nu_\perp+\beta}$ for $x \ll 1$ is the scaling function  \cite{Marro}. So, at the critical point ($\Delta=0$), the density scales as $\overline{\rho}_s\sim L^{-\beta/\nu_\perp}$. Similarly, power laws can be associated to the other quantities at the criticality, namely, $\overline{\zeta}_s\sim L^{-\vartheta/\nu_\perp}$, $\tau\sim L^{\nu_\parallel/\nu_\perp}$, and $\overline{\Omega}_s\sim L^{\gamma/\nu_\perp}$, as shown in FIG. \ref{fig:rhoqs}. Notice that the density of infected nodes $\overline{\Omega}_s$ grows with the node size. The exponents and the critical rates, determined through the null curvature criterion \cite{Ferreira2005}, are listed in Table \ref{tab1}.  These values are, within the margins of error, independent on the number of nodes for $N=3200$, $6400$, and $12800$, or travel rates on the studied interval. In turn, the critical infection rate decreases continuously from $\lambda_c=3.2928$ (the CP critical rate) for $\alpha = 0$ to $\lambda_c =1$ for $\alpha\rightarrow\infty$.

\begin{table}[hbt]
\caption{\label{tab1} Critical rates and exponents for the CP on hybrid networks with $N=1600$ nodes. The uncertainties in the last digits are indicated in parenthesis.}
\begin{tabular}{cccccc}
\hline\hline
$\alpha$ & $\lambda_c$& ~~$\beta/\nu_\perp$~~ & ~~~$\vartheta/\nu_\perp$~~ & ~~~$\gamma/\nu_\perp$~~ & ~~~$\nu_\parallel/\nu_\perp$\\ \hline
0.1 & ~~2.1179(5)~~& 0.526& 0.913& 0.366 & 0.486 \\
0.5 & 1.5435(5)& 0.523 & 0.905& 0.365 &  0.482 \\ 
1.0 & 1.3451(4) & 0.523 & 0.901 & 0.367 & 0.488 \\ \cline{1-2} 
\multicolumn{2}{c}{Mean value} &  0.524(3) & 0.906(6) & 0.366(3) &  0.485(3) \\  \hline\hline
\end{tabular} 
\end{table}

Analogously, for fixed $L$, Eq. (\ref{eq:fss}) leads to similar power laws at the criticality associated to another scaling function $\mathcal{H}(x)$, except for $\overline{\Omega}_s$ that now decays as $N^{\gamma^{\prime}/ \nu_\perp^{\prime}}$ since $\gamma^{\prime}<0$. In Table \ref{tab2}, the critical exponents obtained through distinct approaches to the thermodynamic limit are compared. These values confirm the critical behavior foreseen by Eq. (\ref{eq:fss}) since the sum of the exponents obtained for $L$ and for $N$ fixed equals, within the margins of error, those with a fixed ratio $N/L$. Indeed, $\beta^{\prime \prime} = \nu_\perp^{\prime} \beta + \nu_\perp\beta^{\prime}$,  $\nu_{\perp}^{\prime \prime}=\nuperp \nuperp^{\prime}$ and so forth for $N/L$ fixed. Also, whatever the approach, the relation $\vartheta=\beta +\gamma$ is valid, reflecting the independence between densities of infected individuals inside the nodes and the fraction of cities in which the epidemics persists at criticality. Indeed, as a result, $ \overline{\zeta}_s  \sim \overline{\rho}_s  /  \overline{\Omega}_s  \sim L^{-(\beta+\gamma)/ \nuperp}N^{-(\beta^{\prime}+\gamma^{\prime})/ \nuperp}$.

\begin{table}[hbt]
\caption{\label{tab2} Critical exponents for the densities of the CP on hybrid networks for $\alpha=0.5$ and $\lambda_c = 1.543$. Fixed $L$ FSS analysis were done for $L=1600$ and $3200$ and $N \ge 800$. The ratios $k=0.5$, $1$, and $2$ were used in the $N/L = k$ approach. }
\begin{tabular}{ccccc}
\hline\hline
~~Approach~~  &  \multicolumn{4}{c}{Exponents at $\Delta=0^{a)}$} \\ \cline{2-5} 
              & $\beta/\nu_\perp$~~ & ~~~$\vartheta/\nu_\perp$~~ & ~~~$\gamma/\nu_\perp$~~ &  ~~~$\nu_\parallel/\nu_\perp$\\ \hline
 fixed $N$ & ~$0.526(3)$~ & ~$0.911(6)$~ & ~$0.367(3)$~&~0.484(3)  \\ 
fixed $L$ & ~$0.912(9)$~&~$0.073(8)$~&~$-0.82(1)$~&~0.080(7)  \\
fixed $N/L$  & ~$1.428(6)$~& ~$0.97(1)$~&~$-0.458(8)$~&~0.557(6)  \\ \hline\hline
\end{tabular} \\
$^{a)}$Primes were omitted for sake of brevity.
\end{table}

\begin{figure*}[hbt]
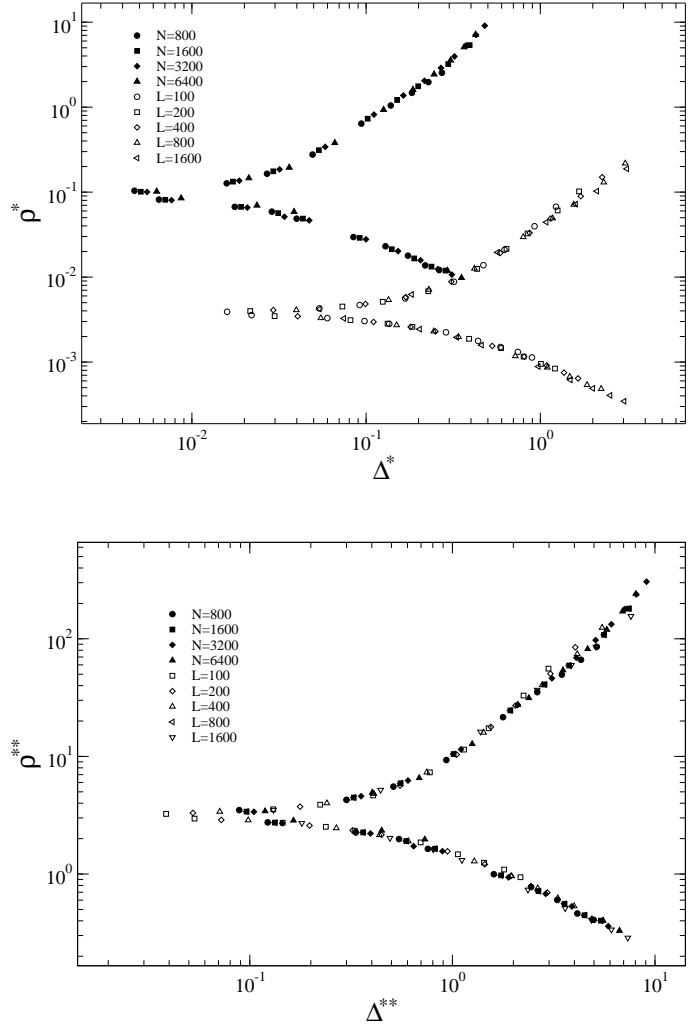

\begin{center}
\includegraphics[width=9cm,height=!,clip=true]{nt_N_L_coll.eps} \\ ~ \\
\includegraphics[width=9cm,height=!,clip=true]{nt_N_L_coll_juntos.eps}
\caption{\label{fig:collapse} Upper: collapses of $\rho$ for either $N=1600$ or $L=800$ fixed. For fixed $N$, $1/\nu_\perp = 0.423$ and $\beta/ \nu_\perp$ taken from Table \ref{tab2} were used. For $L$ fixed, $1/\nuperp^{\prime} = 0.125$ was used. Bottom: collapses for both $L$ and $N$ varying. $\alpha=0.1$ in this Fig.. Here, $\rho^* = \overline{\rho}_s L^{\beta/\nuperp}$ or $\overline{\rho}_s N^{\beta^\prime/\nuperp^\prime}$ and $\Delta^* = \Delta L^{1/\nuperp}$ or $\Delta N^{1/\nuperp^\prime}$. In turn,  $\rho^{**} = \overline{\rho}_s L^{\beta/\nuperp}N^{\beta^\prime/\nuperp^\prime}$ and $\Delta^{**} = \Delta L^{1/\nuperp} N^{1/\nuperp^\prime}$}
\end{center}
\end{figure*}

Finally, according the scaling hypothesis (\ref{eq:fss}), plots of $L^{\beta/\nu_\perp} \overline{\rho}_s(\Delta,L)$ $\times$ $\Delta L^{1/\nu_\perp}$, for fixed $N$, and $N^{\beta^{\prime}/\nu_\perp^{\prime}} \overline{\rho}_s(\Delta,N)$ $\times$ $\Delta N^{1/\nu_\perp^{\prime}}$, for fixed $L$, using the correct exponents $\nu_\perp$ or $\nu_\perp^{\prime}$ should collapse the data onto universal curves. In FIG. \ref{fig:collapse} (upper) are shown the collapses of $\rho$ obtained onto $\mathcal{G}(x)$ (fixed $N$) for $1/\nuperp = 0.423(4)$ and onto $\mathcal{H}(x)$ (fixed $L$) for $1/\nuperp^{\prime} = 0.125(5)$. Thus, using the exponents from Table \ref{tab2}, we found $\beta = 1.23(3)$, $\gamma = 0.86(3)$ and $\nu_\parallel = 1.14(3)$ for fixed $N$, and $\beta^{\prime} = 7.3(3)$, $\gamma^{\prime} = 0.6(3)$ and $\nu_\parallel^{\prime} = 0.6(3)$ for fixed $L$. The large value found for $\beta^{\prime}$, representing a rapid vanishing of $\rho$ at $\lambda_c$, is consistent with the exponential approach obtained for the SIS model in BA networks \cite{Satorras1,Satorras2}. Also, the finite value obtained for $\nu_\parallel$ in the QS analysis for fixed $N$ deserves a comment. It contrasts with the exponential divergence of the characteristic epidemic mean lifetime (FIG. \ref{fig:txnt}). This paradoxical result could be interpreted in terms of two time scales for the infection dynamics. The first one refers to an exponentially long persistence time of the epidemics spreading from a spatially concentrated focus in the subcritical phase. We conjecture that the heterogeneity of the connection among cities, thereby of travels, promotes this long persistence, in contrast to the CP with sources \cite{Marro} if the SF network is replaced by a regular one. The second time scale is associated to the outbreaks, which diverges algebraically at the critical point as in local epidemics. The exponents listed in Tables 1 and 2 are quite distinct from those related to DP universality class \cite{Marro} (fixed $N$) and, also, do not agree with numerical estimates for CP on SF networks \cite{Castellano} (fixed $L$), establishing a new universality class.

\section{Summary}
\label{sec:conclusion}

We presented numerical simulations of the CP on hybrid networks in which the nodes are themselves regular lattices and individuals travel between nodes. The main results are a finite epidemic threshold and an exponentially long persistence of the epidemics below the critical rate, concomitantly with a power law divergence of the outbreak's duration. Generalized scaling functions involving both regular and SF components were proposed and the associated critical exponents determined, demonstrating that the CP on this hybrid network and nonvanishing travel rates establishes a new universality class. Also, at criticality, the densities of infected individuals inside the nodes and the fraction of cities in which the epidemics persists are independent.

\begin{acknowledgments}
This work was partially supported by CNPq and FAPEMIG Brazilian agencies.
\end{acknowledgments}


\begin{thebibliography}{99}
\bibitem{note} A re-emergent disease is one that have occurred in the past at low levels of incidence or isolated cases, but in a given moment returns with high incidence.
\bibitem{r2} The epidemic spreading on regular lattices is defined as $R^2 = (1/n)\langle \sum_i r_i^2 \rangle$, where the sum is done over all occupied sites and $r_i$ is the distance from the original seed where a single infected individual was introduced at $t=0$.
\bibitem{note1} In analogy with FFS analysis in equilibrium systems, in which the critical temperature can be estimated from $T_c(L)=T_c(\infty)+A/L^{1/\nuperp}$.
\end{thebibliography}
\end{document}